\documentclass[10pt]{article}
\usepackage[preprint]{tmlr}
\usepackage{amsmath}
\usepackage[T1]{fontenc}
\usepackage{graphicx}
\usepackage[section]{placeins}
\usepackage{hyperref}
\hypersetup{hidelinks}
\usepackage{url}

\def\TMLRArxiv{1}

\title{Structure is not mechanism: high-gain gated-FFN rows across text and genomic foundation models}

\author{\name Alexandros Tzanakakis \\
      \addr Division of Pharmacology and Toxicology, College of Pharmacy, The University of Texas at Austin, Dell Pediatric Research Institute, Austin, TX, USA \\
      Chandra Family Department of Electrical and Computer Engineering, The University of Texas at Austin, Austin, TX, USA
      \AND
      \name Aris Karatzikos \\
      \addr Division of Pharmacology and Toxicology, College of Pharmacy, The University of Texas at Austin, Dell Pediatric Research Institute, Austin, TX, USA \\
      Department of Computer Science, College of Natural Sciences, The University of Texas at Austin, Austin, TX, USA
      \AND
      \name Ilias Georgakopoulos-Soares \email ilias@austin.utexas.edu \\
      \addr Division of Pharmacology and Toxicology, College of Pharmacy, The University of Texas at Austin, Dell Pediatric Research Institute, Austin, TX, USA}

\begin{document}

\maketitle

{\renewcommand{\thefootnote}{}%
\footnotetext{Large language model tools, including ChatGPT, Claude, and Codex, were used for language editing, code assistance, and LaTeX conversion. All scientific claims, analyses, results, citations, and final text were reviewed and approved by the authors, who take full responsibility for the manuscript.}}

\begin{abstract}
A small number of unusually high-gain parameters can exert disproportionate effects in transformer language models, but whether analogous structures recur in genomic foundation models and whether their structural geometry determines their functional importance remains unknown. We analyzed high-gain rows in gated feed-forward networks across text and genomic foundation models, including a frozen 22-model causal census. By defining an associated bilinear weight operator and computing its magnitude and spectral concentration exactly, without a diagonal approximation, we tested whether structural extremeness constitutes a transferable mechanism. Activation-derived candidates were functionally enriched relative to both random and top-norm same-layer controls, yet neither spectral concentration nor operator magnitude predicted causal effect size, and within the endpoint-homogeneous text-decoder subset these structural associations vanished. A within-layer sweep of 36 rows spanning the full activation-ratio range, performed in one genomic and one text decoder, resolved this enrichment into two regimes: below the detector's acceptance threshold the ratio carried no positive information about causal damage, whereas above it the ratio ordered rows strongly but did not grade their severity as a dose-response. The same sweep revealed that a single layer can contain a second individually catastrophic row that a one-candidate-per-model census cannot detect, and that the damage of co-located critical rows is not additive. Focused mechanistic case studies further showed divergent causal organizations: a robust super-additive pair interaction in DNABERT-2, and in GENERator a sharply position-localized dependence in which preserving or restoring the high-gain row's beginning-of-sequence contribution rescued essentially all native-loss damage. Together these results show that high-gain gated-FFN rows are a recurrent architectural phenotype whose structural prominence acts as an enrichment signal, not a calibrated measure of functional criticality or a specification of causal organization. Enrichment is general, but the mechanism is model-specific.
\end{abstract}

\section{Introduction}

Transformer models often contain small subsets of parameters or activation channels with effects far larger than their numerical prevalence would suggest. In large language models, Yu et al. identified individual ``super weights'' whose ablation can severely degrade generation quality, and related work has described massive activations and attention sinks that recur at fixed positions across prompts \citep{yu2024superweight,sun2024massive,xiao2024efficient,sun2026spike}. These observations raise a broader mechanistic question: are such high-gain structures peculiar to particular language-model families, or are they a more general architectural phenotype that also appears in foundation models trained on other symbolic sequences?

Prior super-weight studies established that individual parameters can be catastrophically important in several text decoders. Here we ask a different question: when activation-derived high-gain structures recur across architectures and domains, which properties of those structures, if any, determine their functional importance? This distinction turns the problem from detecting extreme parameters into testing whether structural extremeness constitutes a transferable mechanism.

Genomic foundation models, such as DNABERT-2, and GENERator, provide a useful setting in which to separate architecture from data domain. Current genomic models span decoder-only generative transformers, bidirectional masked-language encoders, and hybrid sequence models, while operating on nucleotides rather than natural-language corpora \citep{li2026generator,zhou2024dnabert2,boshar2025foundational,brixi2026evo2,zhou2025genomeocean}. If high-gain structures are fundamentally tied to gated feed-forward computation rather than linguistic semantics, related structural phenotypes should recur across text and genomic models. At the same time, their downstream consequences need not be shared; a high-gain coordinate could act as a nearly singular bottleneck in one model, participate in a redundant or interaction-dependent circuit in another, or be structurally conspicuous yet functionally weak.

A central difficulty is therefore definitional. Structural magnitude, spectral concentration, and causal importance are different quantities. A diagonal approximation to the gated-FFN bilinear operator neglects cross terms that can be substantial, particularly in genomic models, making it unsuitable for cross-model dimensionality claims. We therefore define a bilinear weight operator associated with each candidate output row and compute its Frobenius norm and singular spectrum exactly, using these quantities as weight-space descriptors of magnitude and concentration. For multi-row case studies, we evaluate progressively richer response models on held-out finite interventions, asking whether additive predictions suffice or whether calibration and pairwise interaction terms are required \citep{erramilli2026mechanistic}. Candidate rows are suppressed and evaluated through changes in the model's native language-model loss or, for a generative genomic case study, nucleotide composition.

Across the tested panel, high-gain gated-FFN rows form a recurrent structural phenotype across both text and genomic foundation models, but the causal census shows that structural prominence is only an enrichment signal for functional importance, not a calibrated measure of effect size. Across 22 models, frozen activation-selected candidates produced larger native-loss changes than five same-layer controls under both partial suppression and full ablation, yet their effects ranged from negligible or negative changes to several-hundred-percent loss increases. Neither the leading spectral concentration $q_1$ nor operator magnitude predicted the signed full-ablation effect within architecture class, and the highest-norm rows in a candidate's own layer were themselves causally close to inert. We therefore treat singleton functional criticality and multi-component causal response complexity as separate questions. Focused case studies show that DNABERT-2~\citep{zhou2024dnabert2} contains a robust pretrained pair interaction, whereas disruption of a spectrally concentrated GENERator~\citep{li2026generator} row produced a native-loss phenotype mediated almost entirely at the beginning-of-sequence position, alongside a compositional shift that tracked damage magnitude and did not require the row's learned direction. These results support a three-way distinction between structural geometry, functional criticality, and causal response complexity, and argue against treating the super-weight phenomenon as a single universal mechanism.

\section{Results}

\subsection{Structural properties}

We first asked whether activation-derived high-gain gated-FFN rows share a common weight-space structure across foundation models trained on text and genomic sequence. As a retrospective calibration, we applied a diagonal weight-space proxy to the published super-weight coordinates in Llama-7B, Mistral-7B, and OLMo-7B \citep{yu2024superweight,touvron2023llama,jiang2023mistral,groeneveld2024olmo}. In each model, the down-projection output row containing the published super weight ranked first among all output rows in the same layer under this proxy (\textbf{Fig. 1A}). We additionally report the scalar top-1 share, defined as the fraction of the row's diagonal proxy attributable to its single largest hidden-unit contribution. This calibration was used only to connect our row-level analysis to the previously reported scalar super-weight phenomenon. Because the diagonal proxy omits cross terms between hidden units, all subsequent cross-model structural analyses used a row-associated bilinear weight operator $U_k$. Its Frobenius norm quantifies operator magnitude, while its singular spectrum is used to quantify how strongly that magnitude is concentrated in one or more directions.

We quantified spectral concentration using the leading singular-energy fraction $q_1 = \sigma_1^2 / \Sigma_j\sigma_j^2$, where $\sigma_1$ is the largest singular value of $U_k$. Thus, $q_1$ measures the fraction of the operator's total squared magnitude concentrated in its strongest singular direction: $q_1$ = 1 corresponds to a rank-one operator, while values close to 1 indicate a near-rank-one geometry. Across the 23-model structural panel, $q_1$ varied from 0.389 in NTv3~\citep{boshar2025foundational} to 0.9996 in EuroBERT-610M~\citep{boizard2025eurobert} (\textbf{Fig. 1B}). Highly concentrated candidates occurred in both encoders and decoders, whereas several models showed substantially less concentrated structure, indicating that near-rank-one geometry is recurrent but not universal. A regression of $q_1$ on non-embedding parameter count and decoder status explained little of the cross-model variation ($R^2$ = 0.158); neither model size nor decoder status showed convincing evidence of association. Text and genomic models also overlapped substantially in $q_1$. Evo2-7B \citep{brixi2026evo2} was evaluated with the same detector, but its maximum activation ratio was 2.22, below the predefined threshold of 5.0, so no candidate was accepted and no $q_1$ value was assigned.

Because candidate discovery used a single forward pass, we next asked whether the frozen high-gain coordinates remained activation-extreme on independent inputs. Across 20 models for which a comparable singleton candidate could be evaluated, we tested each frozen coordinate on 24 additional inputs without reselecting the row. The candidate remained the highest-activation row in its layer on every tested input in all 20 models and was therefore also within the top 1\% in every case. In five models, the extreme activation was invariant across inputs and localized to a fixed structural position, whereas the remaining 15 showed input-dependent variation in activation magnitude while retaining rank-1 status. Thus, the detected high-gain phenotype was not specific to the original discovery input.

To determine whether this spectral concentration was locally distinctive, we compared each candidate in the 22-model causal-census cohort with five randomly sampled rows from the same layer. For each model, we defined the random-control gap as candidate $q_1$ minus the mean $q_1$ of the five controls. The candidate was more concentrated than all five random controls in every model, and the median random-control gap across models was 0.935 (\textbf{Fig. 1C, top}). Thus, relative to ordinary rows in the same layer, activation-derived candidates appear strongly spectrally exceptional.

This comparison alone, however, does not establish that spectral concentration is distinctive independently of operator magnitude. The random controls carried only 3--8\% of the candidate's Frobenius norm, motivating a stricter comparison with the five highest-Frobenius-norm same-layer rows after excluding the candidate. For this stricter comparison, we defined the top-norm margin as candidate $q_1$ minus the largest $q_1$ among those five neighbours, asking whether even the most spectrally concentrated high-norm alternative remained less concentrated than the candidate. Across the cohort, these controls carried 3.9--133.3\% of the candidate's Frobenius norm (median 21.7\%). In GENERator-PROK-1.2B, two controls exceeded the candidate's norm, making it the only model in which the candidate was not the strict highest-norm row in its layer. The median top-norm margin across models was only 0.014, and in 7 of 22 models at least one high-norm neighbour was more concentrated than the candidate (SmolLM2-360M~\citep{allal2025smollm2}, EuroBERT-210M, EuroBERT-2.1B, ModernBERT-large~\citep{warner2025modernbert}, DNABERT-2, GENERator-EUK-3B, and GenomeOcean-4B; \textbf{Fig. 1C, bottom}). The transferable structural result is therefore narrower: high-gain candidates are unusually large and often highly concentrated, but their spectral concentration is generally not distinctive relative to the strongest available same-layer alternatives.

\begin{figure}[htbp]
\centering
\includegraphics[width=0.95\linewidth]{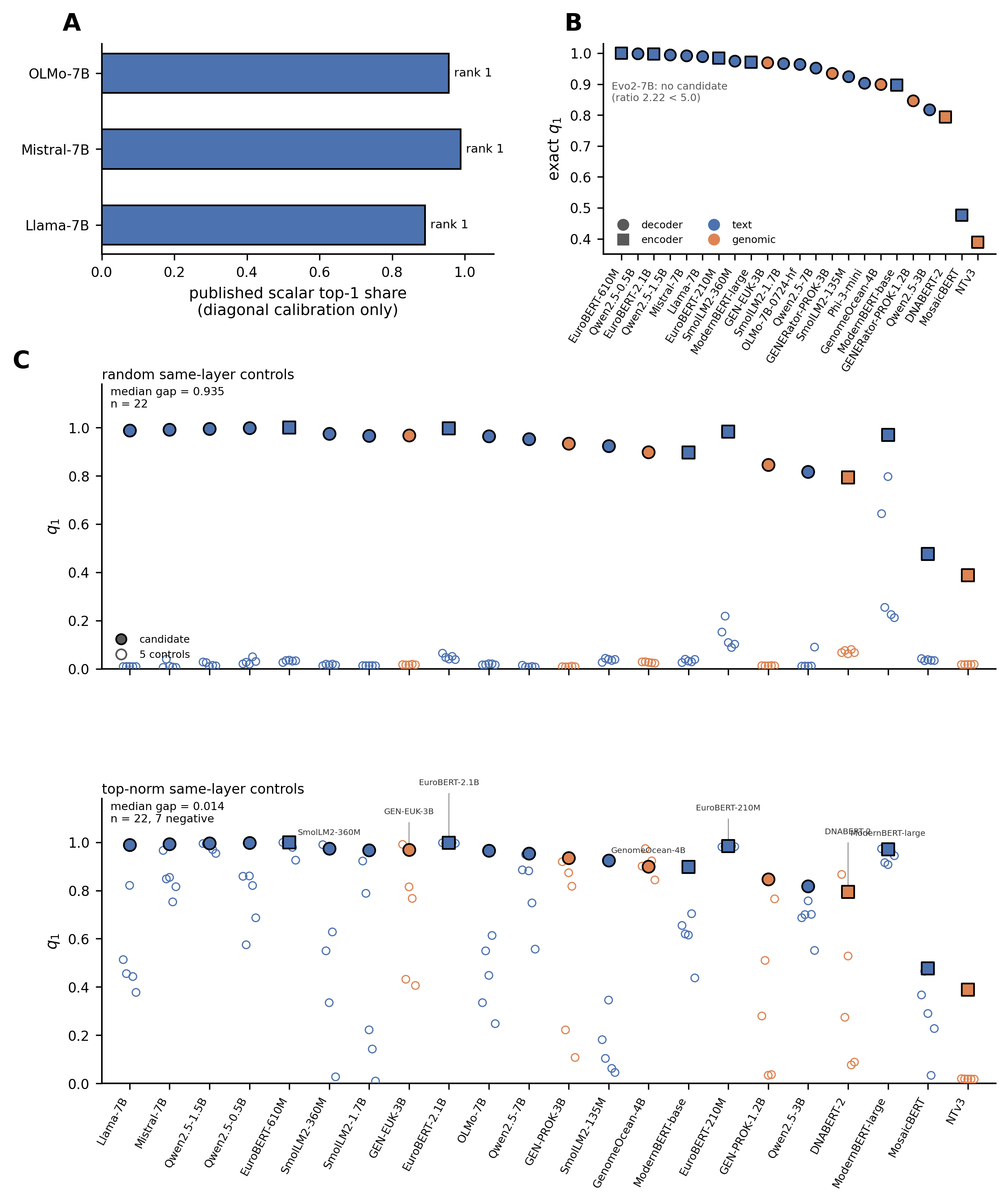}
\caption{\textbf{Structural properties of high-gain gated-FFN rows. (A)} Retrospective diagonal-proxy calibration for the published super-weight-containing rows in Llama-7B, Mistral-7B, and OLMo-7B. \textbf{(B)} Spectral concentration $q_1$ across the 23-model structural panel; the Phi-3 value is the median across its six frozen rows. Evo2-7B did not pass the predefined activation-ratio threshold and is shown without a $q_1$ value. \textbf{(C)} Candidate $q_1$ compared with five random same-layer controls (top) and the five highest-Frobenius-norm rows in the same layer after excluding the candidate in the same layer (bottom). The random-control gap is candidate $q_1$ minus the mean control $q_1$; the top-norm margin is candidate $q_1$ minus the maximum control $q_1$.}
\label{fig:1}
\end{figure}

\subsection{Functional enrichment}

We next asked whether the structurally identified rows were actually important to model function. The singleton causal census comprised 22 models: ten text decoders, six text encoders, four genomic decoders, and two genomic encoders. In each model, one frozen primary candidate and five same-layer control rows were perturbed individually, ensuring that candidate selection was completed before any causal response was observed. We tested two intervention strengths: partial suppression, in which the targeted row was scaled to 50\% of its original value ($\varepsilon$ = 0.5), and full ablation, in which the row was set to zero ($\varepsilon$ = 1.0). Functional effect was measured as the signed relative change in the model's native language-model loss. Positive values therefore indicate that the perturbation worsened the native objective, values near zero indicate little functional effect, and negative values indicate lower loss after perturbation.

At partial suppression, the candidate produced a larger signed native-loss change than the within-model median of the five random same-layer controls in 18 of 22 models (\textbf{Fig. 2A}). The median candidate-minus-control gap across models was +0.60\% (95\% model-bootstrap interval, +0.23\% to +3.06\%), while the median control effect itself was approximately zero. Candidate effects nevertheless varied widely across models, including negative responses in some cases and increases exceeding 300\% in others. Thus, partial suppression showed clear cohort-level enrichment without supporting a universal positive effect.

Full ablation strengthened this separation: the candidate exceeded the median random same-layer control in 20 of 22 models, with a median candidate-minus-control gap of +0.88\% (95\% model-bootstrap interval, +0.61\% to +111.78\%; \textbf{Fig. 2B}, Supplementary Table S3). The strongly asymmetric interval reflects the small, highly right-skewed model-level distribution: most gaps were near zero, whereas a minority were very large, so upper-tail bootstrap resamples could place the median among the large-effect models. Full-ablation candidate effects ranged from $-$0.79\% to +698.87\%, demonstrating that the same structural selection criterion can identify rows whose functional consequences range from negligible or slightly beneficial to catastrophic. The data therefore support functional enrichment, not universal criticality or a common effect size.

\textbf{Magnitude-matched controls}

Random same-layer controls provide only a weak test of specificity because they carry only 3--8\% of the candidate's operator norm. We therefore asked whether the candidate's causal advantage could be explained simply by its unusually large weight-space magnitude.

For each model, we constructed a stricter control set consisting of the five highest-Frobenius-norm rows in the same layer after excluding the candidate. These top-norm controls were selected entirely from frozen weights, without using activations or causal measurements. They therefore represent the strongest alternative rows available to a purely magnitude-based selection rule.

Despite their large norms, the top-norm controls were generally close to functionally inert. Across models, their median relative native-loss effect was +0.035\% under partial suppression and +0.110\% under full ablation. The frozen candidate nevertheless exceeded the within-model median top-norm control in 20 of 22 models at $\varepsilon$ = 0.5 and in 19 of 22 models at $\varepsilon$ = 1.0 \textbf{(Fig. 2D}). Thus, replacing ordinary controls with the largest available same-layer alternatives leaves the candidate's causal enrichment largely intact.

The exceptions are informative but do not overturn the cohort-level pattern. In four models, MosaicBERT at $\varepsilon=0.5$, NTv3 at both intervention strengths, and Qwen2.5-0.5B and Qwen2.5-7B at $\varepsilon=1.0$, the candidate did not exceed the within-model median of the five top-norm controls. NTv3 was the only model in which the candidate underperformed both the random-control and top-norm-control medians at both strengths. The main result is therefore not that activation-derived candidates are uniquely causal in every model, but that activation-based selection generally identifies functionally consequential rows that simple weight-magnitude ranking does not recover.

We could not identify a structural property that explained the minority of models in which a top-norm control was causally competitive with the candidate. To test whether local spectral geometry distinguished these cases, we defined the absolute top-norm structural margin as |candidate $q_1$ $-$ max $q_1$ among the five top-norm neighbours|. This quantity was not significantly associated with the candidate-minus-top-norm-control causal gap at either intervention strength (Spearman $\rho$ = 0.311, p = 0.159 at $\varepsilon$ = 0.5; $\rho$ = 0.355, p = 0.105 at $\varepsilon$ = 1.0). We therefore treat these exceptions as an unexplained minority rather than as a structurally predictable regime.

More importantly, several models showed a direct dissociation between structural geometry and causal importance. In DNABERT-2 and GENERator-EUK-3B, the frozen candidate was less spectrally concentrated than at least one of the five highest-norm rows in the same layer, yet its ablation caused one to three orders of magnitude more native-loss damage. Thus, weight-space concentration and causal importance can rank nearby rows in opposite orders.

Across the full 22-model cohort, candidate $q_1$ was essentially unrelated to the signed full-ablation effect (Spearman $\rho$ = 0.074, p = 0.744; 95\% model-bootstrap interval, $-$0.391 to 0.513; \textbf{Fig. 2C}, Supplementary Table S4). Highly concentrated candidates could therefore be either weakly or catastrophically causal. Layer-relative Frobenius magnitude showed a nominal association with full-ablation effect across the heterogeneous panel ($\rho$ = 0.441, p = 0.040), but this relationship was weaker and non-significant under partial suppression ($\rho$ = 0.348, p = 0.112), and no multiple-comparison correction was applied across predictors and intervention strengths. Moreover, within the endpoint-homogeneous text-decoder subset, both associations disappeared and changed sign at full ablation (Frobenius $\rho$ = $-$0.042, p = 0.907; $q_1$ $\rho$ = $-$0.333, p = 0.347). Together with the near-inert behavior of the top-norm controls, these results argue against either spectral concentration or operator magnitude as a robust, transferable predictor of causal effect size. They do not exclude a nonlinear threshold effect specific to the most extreme high-gain row.

Taken together, the singleton census narrows the relationship between structure and function. Activation-derived high-gain rows are usually more functionally consequential than both ordinary and high-norm same-layer controls, showing that structural selection enriches for functional importance. However, neither spectral concentration nor operator magnitude provides a reliable measure of how large that causal effect will be. Structural geometry can therefore help identify unusual candidate locations, but functional severity must be established by intervention.

The census deliberately tests one primary row at a time and therefore cannot determine how multiple high-gain components combine. In particular, a row with a small singleton effect could still participate in a strongly interaction-dependent system. We therefore treat singleton functional criticality and multi-component causal response complexity as separate questions and examine the latter only in focused mechanistic case studies.

\begin{figure}[htbp]
\centering
\includegraphics[width=0.95\linewidth]{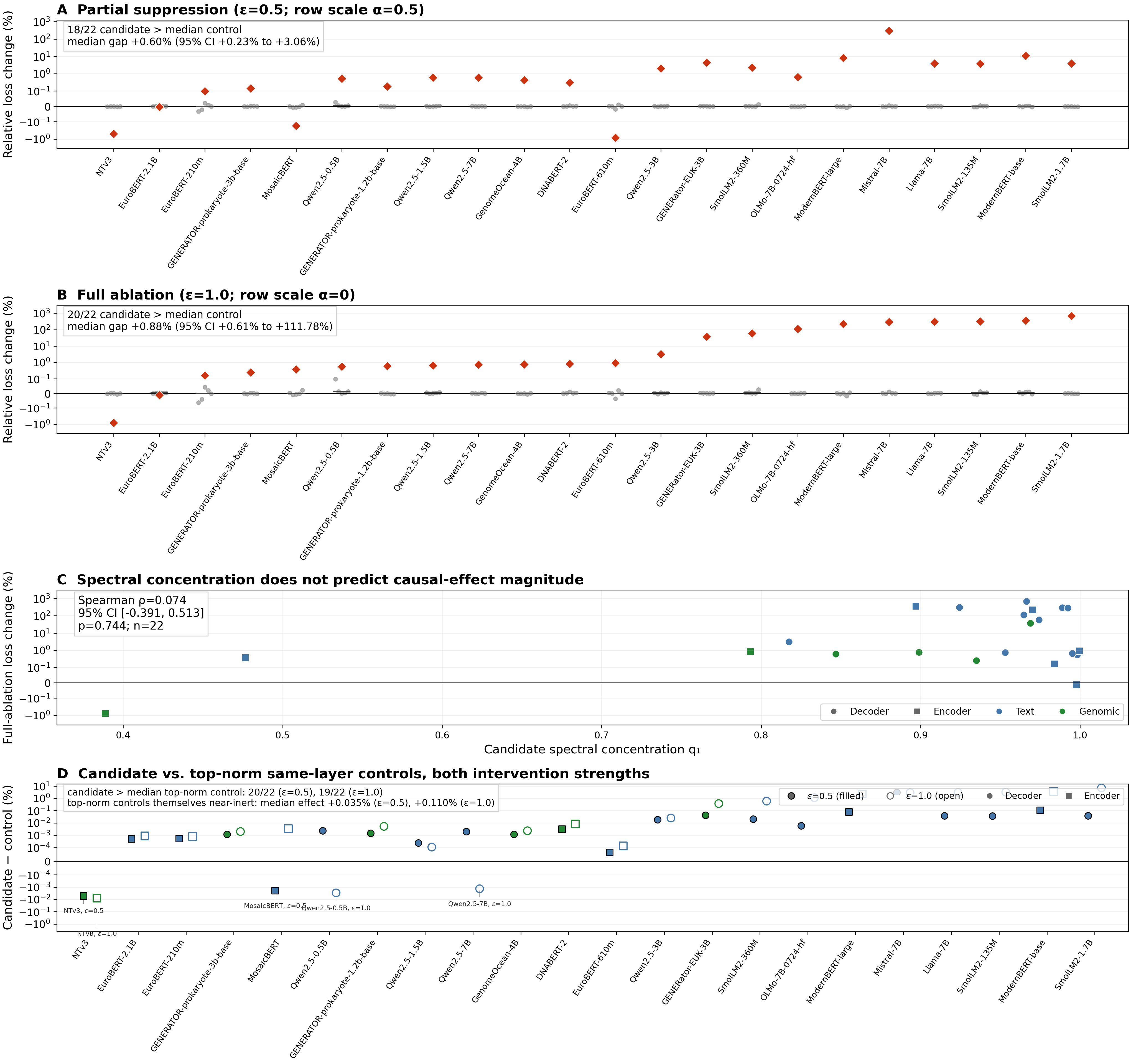}
\caption{\textbf{Causal effects of frozen high-gain candidates across the 22-model census. (A)} Signed relative native-loss change after partial suppression ($\varepsilon$ = 0.5) for each candidate and five random same-layer controls. \textbf{(B)} Corresponding effects under full ablation ($\varepsilon$ = 1.0). Diamonds denote candidates, grey points individual controls, and horizontal bars within-model control medians. \textbf{(C)} Candidate spectral concentration $q_1$ versus signed full-ablation native-loss effect. \textbf{(D)} Candidate-minus-median-top-norm-control causal gap at both intervention strengths. Symmetric-log axes are used where necessary to display negative, near-zero, and large positive effects on the same scale.}
\label{fig:2}
\end{figure}

\subsection{Within-layer sweep: detection versus severity}

The comparisons above are either between models, where each model contributes a single candidate and model identity is therefore confounded with that candidate's activation ratio, or within models, where the candidate is compared against five controls that sit near ratio 1 by construction. Neither design can establish whether causal damage scales with the ratio. We therefore measured both quantities for many rows inside a single layer, holding architecture, scale, evaluation endpoint, and baseline loss fixed.

In each of two decoders, GENERator-EUK-3B at layer 4 and SmolLM2-1.7B at layer 7, the layers carrying their frozen candidates r2371 and r227, we recorded the maximum absolute down-projection output activation of every row, formed each row's layer-relative activation ratio, and selected 36 rows log-spaced by ratio rank so that the full range was represented rather than only its extremes. Selection used the activation ratio alone and never a causal outcome. Each selected row was then ablated ($\alpha$ = 0) and the model's native loss re-evaluated on its own frozen evaluation pool.

Pooled across all 36 rows, ratio and relative native-loss increase were positively associated in both models (GENERator $\rho$ = 0.454, p = 0.005; SmolLM2 $\rho$ = 0.625, p $<$ 0.001), and the association survived removal of the frozen candidate itself ($\rho$ = 0.406 and $\rho$ = 0.592), so it does not depend on a single extreme point. The pooled coefficient is nevertheless misleading, because the relationship is not uniform across the range. Partitioning the swept rows at the detector's own acceptance threshold of 5.0 separates two regimes. Among the 21 rows below threshold the ratio carried no positive information about damage in either model: the association was indistinguishable from zero in GENERator ($\rho$ = 0.040, p = 0.862) and negative in SmolLM2 ($\rho$ = $-$0.584, p = 0.005) but at relative-damage magnitudes of order $10^{-5}$, at which the sign of the association is not physically interpretable. Below threshold the ratio therefore carries no usable information about damage in either model. Among the 15 rows at or above threshold the ratio ordered rows strongly (GENERator $\rho$ = 0.639, p = 0.010; SmolLM2 $\rho$ = 0.975, p $<$ 0.001) and again survived removal of the candidate ($\rho$ = 0.556 and $\rho$ = 0.969). The frozen candidate's damage exceeded that of the median sub-threshold row by factors of $6.1 \times 10^{4}$ and $5.4 \times 10^{5}$ (\textbf{Fig. 3, Supplementary Table S7}).

We report this partition as exploratory. The split point is the detector's acceptance threshold, which was fixed before these measurements and not fitted to them, but the decision to analyze the two regimes separately followed inspection of the swept data. The consistent reading across both domains is that the activation ratio acts as a detector and, above its own threshold, as an ordering, but not as a dose-response measure of severity. This is the within-model counterpart of the cohort-level dissociation reported above, and it supports rather than revises the description of structural prominence as an enrichment signal.

\begin{figure}[htbp]
\centering
\includegraphics[width=0.95\linewidth]{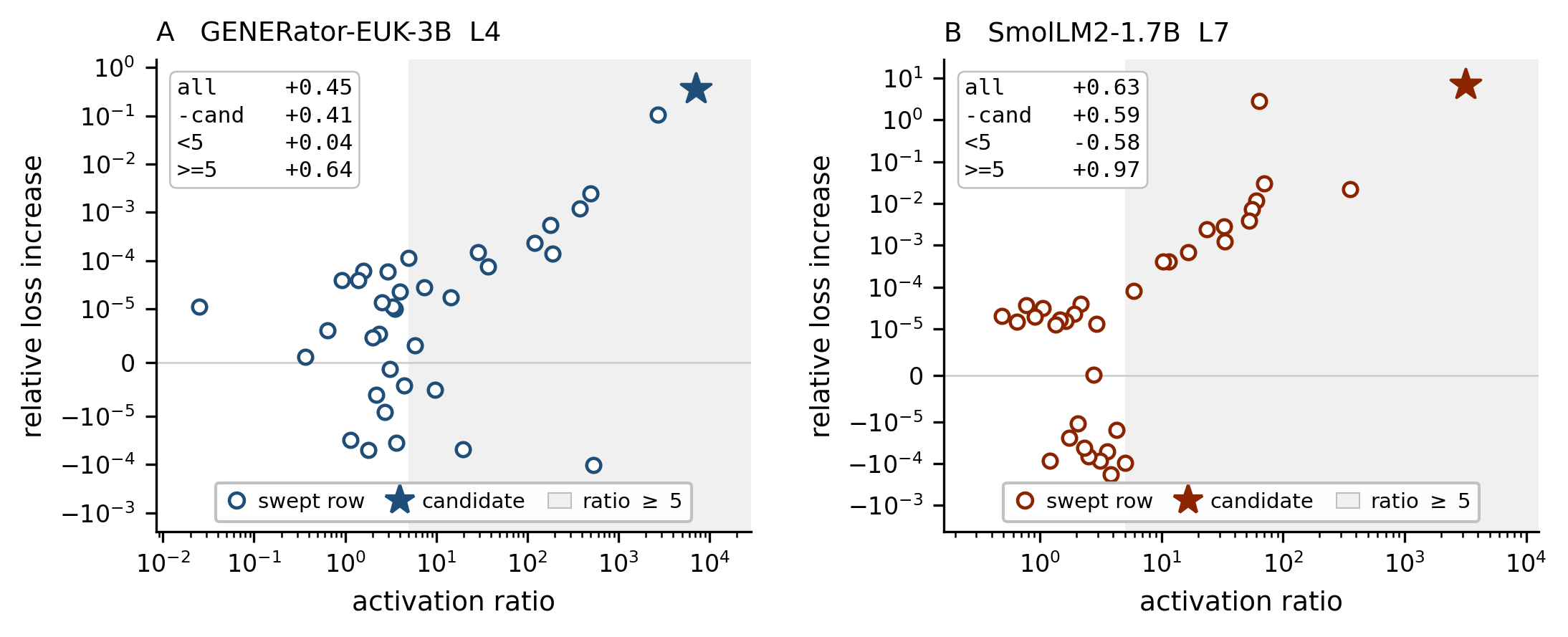}
\caption{\textbf{Activation ratio and single-row ablation effects across rows within a layer.} Damage from deleting a single row (y-axis) is plotted against that row's activation ratio (x-axis) for 36 rows sampled within one layer of the following models. \textbf{(A)} GENERator-EUK-3B, layer 4, and \textbf{(B)} SmolLM2-1.7B, layer 7. Stars indicate each model's main candidate row. The shaded region denotes rows satisfying the activation-ratio threshold of 5.}
\label{fig:3}
\end{figure}

\subsection{Two critical rows in one layer}

Sweeping many rows rather than one also exposed structure that a one-candidate-per-model census cannot detect. In SmolLM2-1.7B layer 7, two rows were individually catastrophic under ablation: the frozen candidate r227 (activation ratio 3181.7, rank 1 of 2048; relative native-loss increase +698.9\%) and r161 (ratio 63.5, rank 4; +287.2\%). A row where the detector ranks above r161 was close to inert: r749 (ratio 358.6, rank 2) cost +2.2\%. A row ranked fourth by activation ratio was therefore 130-fold more damaging than the row ranked second, and r161 appears in no census artifact because the census retains one candidate per model.

Ablating rows in pairs on the same evaluation pool showed that these effects are not independent. Because the endpoint is a loss increase, a positive interaction term denotes super-additivity. Joint ablation of r227 and r161 cost +866.6\% against a sum of singles of +986.1\% (interaction $-$119.5 percentage points, or 87.9\% of the additive expectation), which is mildly sub-additive, whereas joint ablation of r227 and r749 was super-additive (+772.9\% against +701.1\%). The third pair behaved qualitatively differently: ablating r161 together with r749 cost +2.1\% against a sum of singles of +289.4\%, or 0.7\% of the additive expectation. Removing r749, which is nearly harmless alone, therefore abolishes the catastrophic effect of removing r161, and removing both leaves the model essentially intact. Because a cancellation of this form is also the signature of a weight-restoration error, all six conditions were re-measured with an independent implementation that zeroes rows by explicit indexing, restores them from separately held copies, and re-evaluates the intact loss after every restoration; every value reproduced exactly, the result was invariant to ablation order, and the recovered intact loss was identical to ten significant figures (Supplementary Table S8).

The three rows are also the geometric extremes of their layer. Referenced against all 19,900 pairs among the 200 highest-norm rows of the layer (mean cosine $-$0.0001, s.d. 0.0224), r161 and r749 are the most aligned pair in the layer (cosine +0.395, 100th percentile) and r227 and r749 the most anti-aligned (cosine $-$0.398, 0th percentile), while r227 and r161 are unremarkable (cosine $-$0.032). The two rows whose ablations cancel are therefore positively aligned rather than opposed, which excludes the simplest mutual-cancellation account. Their operator norms differ approximately threefold. We report the causal and geometric observations as measurements and do not propose a mechanism linking them, since establishing why an aligned, higher-norm partner is required for the damage of r161 to appear would require activation-level mediation experiments that we did not perform.

These two multi-row studies do not reduce to a single interaction sign. In DNABERT-2, two individually weak rows combine super-additively. In SmolLM2-1.7B layer 7 the sign varies from pair to pair within that one layer: r227 with r749 is likewise super-additive, r227 with r161 is mildly sub-additive, and r161 with r749 cancels almost entirely. Interaction between high-gain rows is therefore substantial in both models, but its sign is not a property of the model or even of the layer which argues against a single universal interaction mechanism.

\subsection{DNABERT-2: a pairwise interaction}

The singleton census cannot reveal interactions between multiple high-gain components, so we next asked whether rows with weak individual effects could nevertheless participate in a strongly non-additive system. In DNABERT-2, we analyzed a fixed 10-row high-gain basis on the pretrained masked-language-model objective using 256 hg38~\citep{schneider2017grch38} windows and a fixed masking realization. Two rows, L9/r264 and L9/r294, had only small individual effects under full ablation, increasing loss by +0.0218 and +0.0064, respectively. When both rows were ablated together, however, loss increased by +2.0400. We quantified this non-additivity as pairwise epistasis, defined as joint effect minus the sum of the two singleton effects: $E_{AB} = d_{AB} - d_A - d_B$. The resulting epistasis was +2.0118 under full ablation and remained positive (+0.0528) under partial suppression ($\varepsilon$ = 0.5; \textbf{Fig. 4A,B}). Thus, two rows that appear nearly unimportant when tested separately can become strongly consequential when perturbed together.

To determine whether this non-additivity extended beyond a single exceptional pair, we applied the observer-family progression motivated by the mechanistic tomography framework \citep{erramilli2026mechanistic} to the fixed 10-row DNABERT-2 system. Each observer family represented a distinct hypothesis about how row effects combine. F0 assumed naive additivity, predicting the response to a multi-row intervention by summing the independently measured singleton effects. F1 asked whether this additive prediction had the correct structure but the wrong overall scale, and therefore fitted a single global calibration factor. F2 allowed each row to have its own fitted contribution while remaining strictly additive. Finally, F3 added all 45 pairwise interaction terms, allowing the effect of perturbing one row to depend on whether another row was perturbed simultaneously. The models were fitted and calibrated on separate intervention subsets and evaluated on 20 held-out conditions at each intervention strength. Thus, improved held-out performance from F2 to F3 specifically tests whether pairwise interaction terms provide predictive information beyond a flexible additive model.

Pairwise terms substantially improved held-out prediction at both intervention strengths (\textbf{Fig. 4C, Supplementary Table S6}). At $\varepsilon$ = 0.5, held-out $R^2$ increased from $-$0.023 for naive additivity (F0) and 0.517 for the fitted additive model (F2) to 0.888 for the pairwise model (F3); relative to F2, F3 reduced held-out MAE by 54.7\% (95\% CI, 15.0\% to 56.1\%). Under full ablation, F3 achieved a held-out $R^2$ = 0.790 compared with 0.581 for F2 and reduced MAE by 24.4\% (95\% CI, 17.2\% to 28.8\%). Moreover, F3 outperformed F2 on held-out MAE in all 100 alternative fit/calibration/held-out resplits at both intervention strengths, showing that the advantage of pairwise terms was not specific to one partition of the measured conditions.

The directly measured L9/r264--L9/r294 interaction also re-emerged without being privileged during model fitting. Among all 45 pairwise F3 coefficients, this pair ranked first by absolute coefficient at both $\varepsilon$ = 0.5 and $\varepsilon$ = 1.0 (\textbf{Fig. 4D}). Thus, the same interaction identified by direct joint ablation was independently recovered as the dominant pairwise term in the held-out tomography analysis. DNABERT-2 therefore provides a robust example of a high-gain system whose finite-intervention response is genuinely interaction-dependent on the pretrained masked-language-model objective.

\begin{figure}[htbp]
\centering
\includegraphics[width=0.95\linewidth]{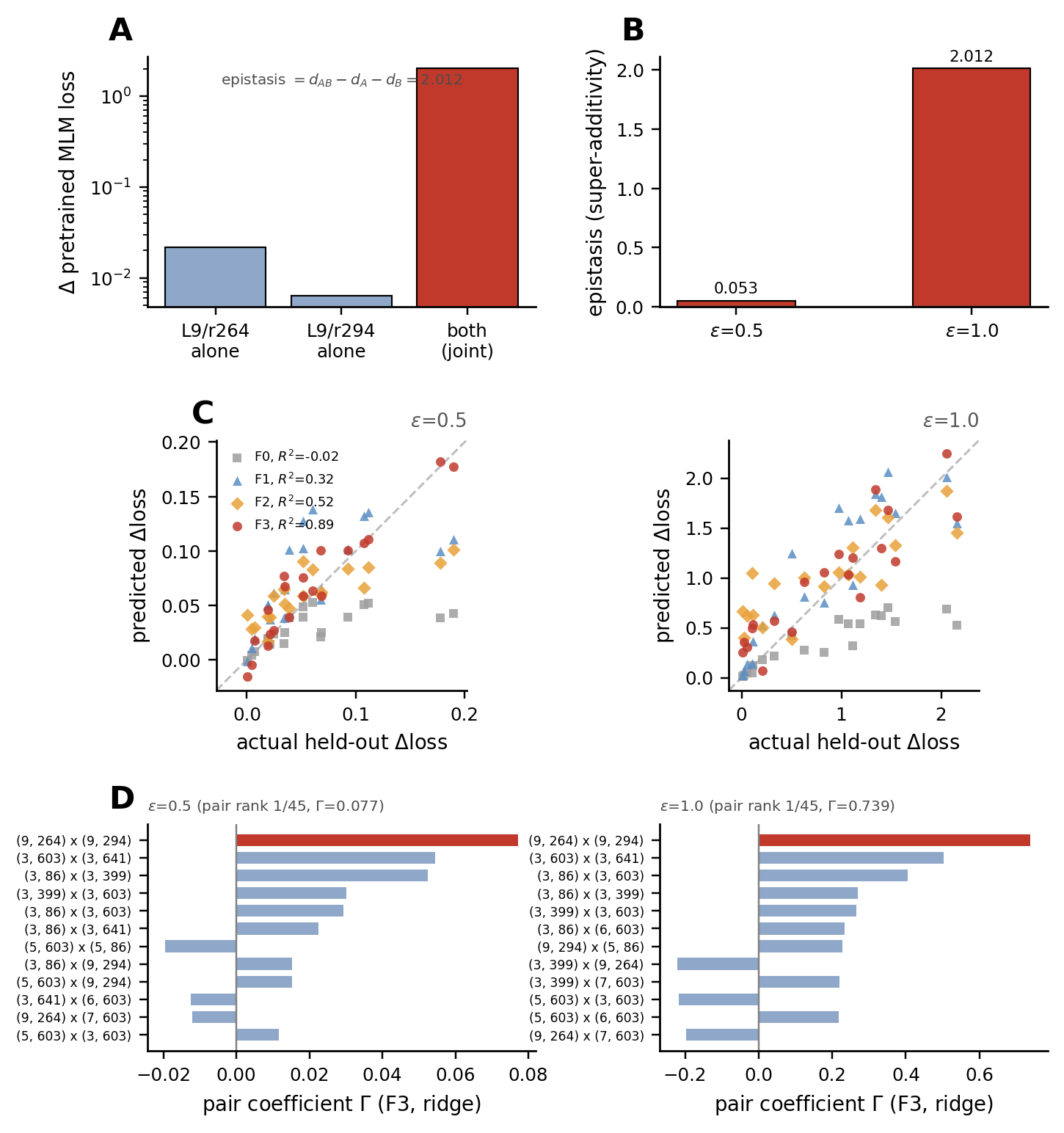}
\caption{\textbf{DNABERT-2 multi-row causal-response analysis. (A)} Full-ablation effects of L9/r264 and L9/r294 individually and jointly. \textbf{(B)} Pairwise epistasis at $\varepsilon$ = 0.5 and $\varepsilon$ = 1.0, defined as the joint effect minus the sum of the singleton effects; these measurements come from a standalone direct pair-ablation experiment. \textbf{(C)} Held-out predictive performance of observer families F0--F3 across multi-row intervention conditions. \textbf{(D)} Fitted F3 pairwise interaction coefficients across all 45 row pairs.}
\label{fig:4}
\end{figure}

\subsection{GENERator: position-localized dependence}

GENERator EUK provides a complementary mechanistic case study focused on a single high-gain location. Its primary candidate, L4/r2371, has $q_1$ = 0.969, indicating that its associated bilinear weight operator is strongly concentrated in one singular direction (\textbf{Fig. 5A}). As contextual characterization, a separate reproduction also revealed a strong beginning-of-sequence (BOS) phenotype: position 0 received 37.9\% of mean incoming attention, corresponding to 33.0 times the uniform expectation, and 78.8\% of layer-head observations had their maximum attention at position 0. The maximum high-gain activation was also observed at position 0 (\textbf{Fig. 5B}). These measurements established that the high-gain activation co-localized with a strongly privileged beginning-of-sequence position. We therefore asked whether the functional effect of row 2371 was itself concentrated at this position. Because attention was not independently manipulated, the experiments below test positional mediation of the row's functional effect but do not establish that row 2371 causes the attention sink.

Position-specific intervention revealed that essentially the entire native-loss effect of row 2371 was mediated through its contribution at the beginning-of-sequence position (\textbf{Fig. 5C}). The intact model had a native NLL of 6.385, whereas full row ablation increased NLL to 8.754. Ablating the row contribution only at BOS produced NLL = 8.722, reproducing nearly all of the full-ablation damage. Conversely, ablating the row everywhere except BOS yielded NLL = 6.386, corresponding to a rescue fraction of 0.9996 (95\% CI, 0.998--1.002). Full row ablation followed by restoration of the intact BOS contribution likewise restored NLL to 6.386, whereas restoration at a matched non-BOS position left NLL at 8.754 and produced essentially no rescue. Thus, for this coordinate, loss of the BOS contribution was approximately sufficient for the full native-loss phenotype, while preservation or restoration of that contribution was approximately sufficient to prevent it.

The positional dependence extended to the generated-sequence composition phenotype (\textbf{Fig. 5D}). Full row ablation reduced mean GC fraction from approximately 0.420 in the intact model to 0.307. Preserving only the BOS contribution or restoring that contribution after full ablation raised GC fraction to approximately 0.412, whereas restoration at the matched non-BOS position left it near the ablated value (0.307). BOS-only ablation reduced GC fraction further to 0.292 rather than simply reproducing the full-ablation value, indicating that the compositional response is damage-linked but not fully determined by native-loss magnitude. The random-direction experiment shows the GC shift does not require the row's learned direction and scales with damage, while the BOS-only condition shows that the position at which damage is inflicted further modulates the compositional outcome beyond its native-loss magnitude. The two observations are consistent. Composition tracks damage, but not damage magnitude alone. Nevertheless, the preservation and restoration experiments show that the large GC shift, like the native-loss phenotype, depends strongly on the row's contribution at BOS.

The perturbation was not specific to GC composition. Row-2371 ablation also reduced predictive entropy by 0.50 nats, modestly reduced higher-order sequence diversity, and increased short-range repetitiveness, while the non-ACGT filtering rate remained exactly zero. Thus, the model continued to generate valid nucleotide sequence rather than arbitrary or malformed output, but the GC shift occurred alongside a broader, milder degradation of generation quality. We therefore interpret GC fraction as the most direct compositional readout of the phenotype rather than as an isolated affected property.

A complementary perturbation tested whether the compositional phenotype required the learned direction of row 2371. Five non-candidate rows remained comparatively insensitive and could not be damage-matched to row-2371 ablation. We therefore replaced row 2371 with one fixed random unit direction and varied its scale. At a scale producing essentially the same native-loss damage as full ablation, the random replacement also reproduced the low-GC phenotype and associated changes in entropy, diversity, repetitiveness, and sequence validity. Across the scale sweep, GC composition recovered as native-loss damage decreased, although even the largest random replacement remained severely damaging (\textbf{Supplementary Table S5}). These results therefore support damage-linked composition at this load-bearing location, while the position-specific interventions above further identify BOS as the location through which the row's native-loss effect is predominantly expressed. Detailed random-direction results are provided in Supplementary Fig. S2.

The compositional shift was also not dominated by a single trivial sequence artifact. The longest homopolymer accounted for only 3.9\% of the GC decrease, non-ACGT output remained absent, and enriched short motifs after ablation were broadly AT-rich. Together with the broader but milder changes in entropy and sequence diversity, these results indicate a genuine damage-associated shift in generated sequence composition rather than collapse to a single repetitive or invalid output mode.

Together, the DNABERT-2 and GENERator case studies illustrate distinct causal organizations underlying high-gain structure. DNABERT-2 exhibits strong pairwise dependence on its pretrained masked-language-model objective, such that individually weak rows become consequential when perturbed jointly. GENERator instead exhibits sharply position-localized dependence: essentially the entire native-loss effect of row L4/r2371 is mediated through its contribution at BOS, with restoration at a matched non-BOS position providing no rescue. The complementary random-direction experiment further shows that the associated compositional phenotype does not require preservation of the learned row direction under comparably severe damage. These qualitatively different mechanisms reinforce the central conclusion that structural prominence does not specify a universal causal organization.

\begin{figure}[htbp]
\centering
\includegraphics[width=0.95\linewidth]{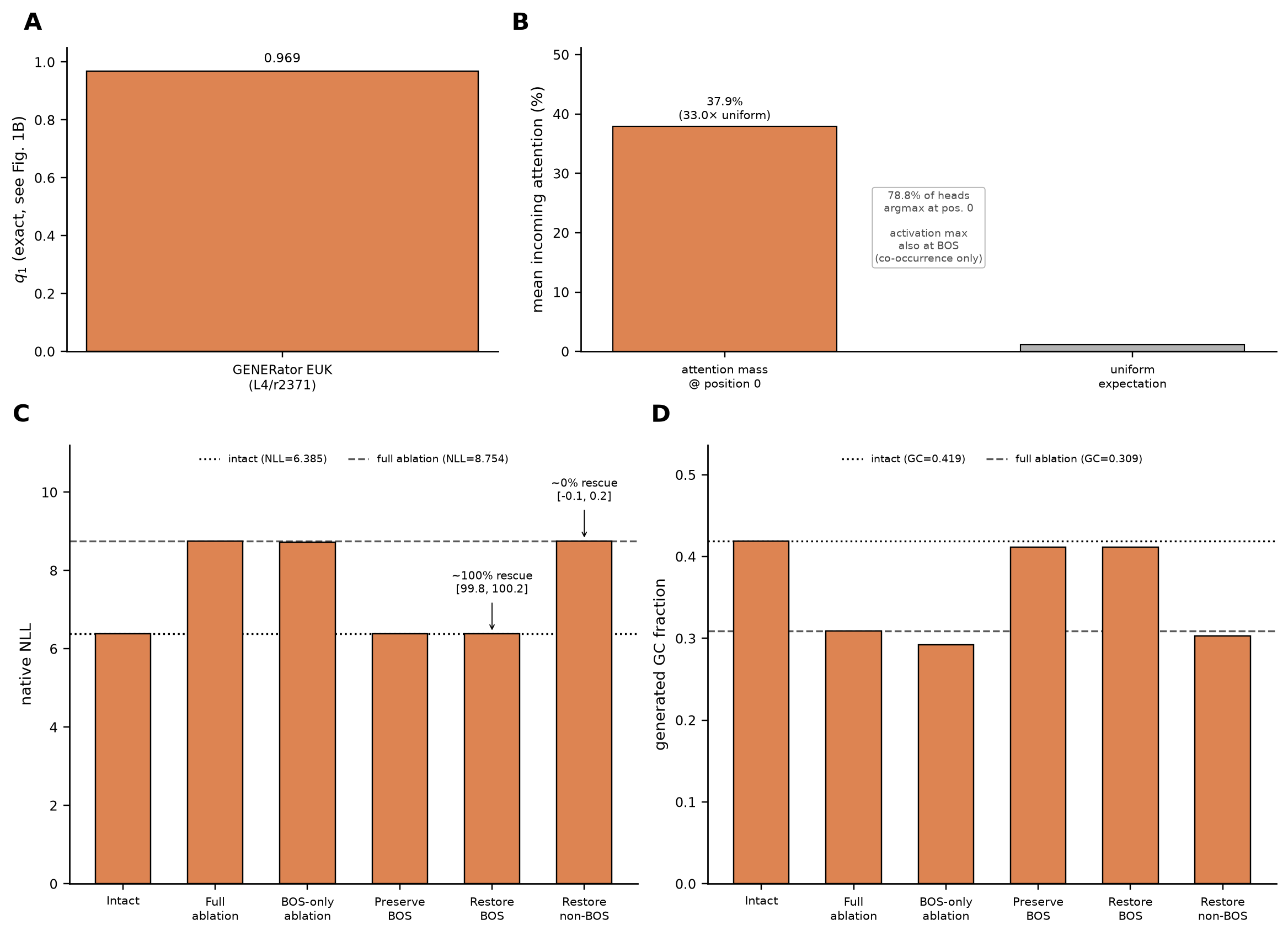}
\caption{\textbf{Beginning-of-sequence mediation of GENERator EUK row L4/r2371. (A)} Spectral concentration $q_1$ of row L4/r2371. \textbf{(B)} Beginning-of-sequence attention and activation measurements at position 0; these measurements establish co-localization but do not test whether the row causes the attention sink. \textbf{(C)} Native language-model loss under intact, full-ablation, BOS-only ablation, BOS-preservation, BOS-restoration, and matched non-BOS-restoration interventions. Removing the row contribution only at BOS reproduced nearly all full-ablation damage, whereas preserving or restoring the BOS contribution produced approximately complete rescue; restoration at the matched non-BOS position produced essentially none. \textbf{(D)} Generated GC fraction under the same position-specific interventions. Error intervals denote paired 95\% bootstrap intervals.}
\label{fig:5}
\end{figure}

\section{Discussion}

Our study separates three questions that are easily conflated across examinations of super-weight-like phenomena: whether a component is structurally unusual, whether it is functionally critical when perturbed, and how multiple components combine under finite intervention. Across text and genomic foundation models, we find that activation-derived high-gain gated-FFN rows formed a recurrent structural phenotype characterized by unusually large row-associated bilinear weight operators and, often, strong spectral concentration. These properties were useful for identifying exceptional locations, but they did not calibrate causal severity. Activation-derived candidates were generally more consequential than both ordinary and high-norm same-layer controls, yet structurally similar rows could have very different ablation effects. Therefore, we conclude that structural geometry is an enrichment signal for functional importance, not a direct readout of mechanism.

This narrows the interpretation of the original super-weight phenomenon rather than contradicting it. Previous work established that individual scalar weights can be catastrophically important in several text decoders \citep{yu2024superweight}. Our results show that the broader recurrence of high-gain structure should not be interpreted as evidence for a universal causal class with a common effect size or mechanism. The more transferable object is the structural phenotype itself; catastrophic singleton dependence is one possible functional realization of that phenotype. Conversely, simple weight-magnitude ranking does not recover the same causal ordering as activation-derived selection, indicating that structural magnitude alone is insufficient even for identifying which nearby rows will matter most.

The within-layer sweep bounds how the activation-ratio criterion should be described. Inside a single layer the ratio is informative about which rows are candidates and, above its acceptance threshold, about their relative ordering, but it is uninformative below that threshold and it does not recover the most consequential coordinate. A row ranked fourth by ratio was 130-fold more damaging than the row ranked second in SmolLM2-1.7B, and in OLMo-7B the coordinate that the ratio selects is 47-fold less damaging than the published one that it does not. The criterion is therefore best specified as a reproducible detector for high-gain candidates rather than as a definition of the functionally critical row.

The focused case studies further separate functional criticality from causal response complexity. In DNABERT-2, rows with weak singleton effects produced strong pairwise interactions that were supported both by direct joint intervention and by improved held-out prediction when pairwise terms were added to the observer model. The relevant mechanism is therefore interaction-dependent on the pretrained masked-language-model objective rather than reducible to the importance of either row in isolation. GENERator showed a different organization. Position-specific intervention demonstrated that essentially the entire native-loss effect of row L4/r2371 was mediated through its contribution at BOS: removing only that contribution reproduced the full-ablation phenotype, whereas preserving or restoring it produced nearly complete rescue, and restoration at a matched non-BOS position did not. The accompanying GC phenotype was also largely rescued by BOS preservation or restoration, while the random-direction experiment showed that matched severe damage did not require preservation of the learned row direction. Thus, the supported mechanism is a sharply position-localized dependence at a load-bearing BOS pathway rather than direction-specific control of GC composition. The co-localized attention-sink phenotype remains observational because attention itself was not causally manipulated.

Several limitations bound these conclusions. The model panel is heterogeneous and is not a random sample from a defined population; architectures, tokenizers, training domains, and native evaluation objectives differ across models. The cohort census tests one frozen primary row per model and therefore measures singleton functional criticality rather than the interaction structure of complete high-gain systems. The within-layer sweep partially offsets this limitation for one model, and shows that the restriction is consequential. The swept layer contained a second individually catastrophic row that the census design does not report. Full ablation is also a strong intervention that can enter nonlinear failure regimes. The GENERator BOS-mediation result concerns a single model and a single high-gain coordinate and therefore should not be assumed to generalize to other genomic decoders or high-gain rows. The GENERator specificity experiment used one fixed random direction, and neither that replacement nor the non-candidate control rows produced a condition in which the learned direction was absent while normal function was restored. Finally, mechanistic claims here concern pretrained native objectives rather than downstream biological or linguistic tasks, and the recorded checkpoint provenance remains incomplete for NTv3.

The evidence in this study sits at three different scales, and we are explicit about which claims each supports. The enrichment of activation-derived candidates over same-layer controls, and the failure of weight-space geometry to predict causal severity, are cohort-scale results established across 22 models. The dissociation between ordering and severity is a within-layer result demonstrated in two decoders. The multi-row interaction, the beginning-of-sequence mediation, and the two-critical-rows structure are single-model existence proofs, not population claims. We therefore do not assert that any specific causal organization generalizes across models. The organizing claim of this work is precisely the opposite, that enrichment is a general property of high-gain structure whereas its mechanistic realization is model-specific, and the case studies are worked examples of that heterogeneity rather than instances of a shared mechanism.

The broader implication is methodological. Weight-space analysis can identify recurrent structural phenotypes and prioritize candidate components, but mechanisms must be established through intervention. Functional criticality should therefore be measured rather than inferred from operator magnitude or spectral concentration, and multi-component causal response complexity should be tested separately from singleton effects. Held-out observer comparisons provide one way to make that second step explicit without assuming in advance that richer interaction models are necessary. This separation of structural geometry, functional criticality, and causal response complexity provides a more conservative framework for interpreting extreme parameter pathways across foundation models.

\section{Methods}

\subsection{Model panel and checkpoints}

We analyzed gated feed-forward structures across 23 foundation models with accepted activation-based candidates. The text-decoder panel comprised Llama-7B~\citep{touvron2023llama}, Mistral-7B~\citep{jiang2023mistral}, OLMo-7B~\citep{groeneveld2024olmo}, Phi-3-mini~\citep{abdin2024phi3}, Qwen2.5-7B, Qwen2.5-0.5B, Qwen2.5-1.5B, Qwen2.5-3B~\citep{yang2024qwen25}, SmolLM2-135M, SmolLM2-360M, and SmolLM2-1.7B~\citep{allal2025smollm2}. Text encoders comprised MosaicBERT~\citep{portes2023mosaicbert}, ModernBERT-base, ModernBERT-large~\citep{warner2025modernbert}, EuroBERT-210M, EuroBERT-610M, and EuroBERT-2.1B~\citep{boizard2025eurobert}. Genomic decoders comprised GENERator-v2-eukaryote-3B, GENERator-v2-prokaryote-1.2B, GENERator-v2-prokaryote-3B~\citep{li2026generator}, and GenomeOcean-4B~\citep{zhou2025genomeocean}; genomic encoders comprised DNABERT-2~\citep{zhou2024dnabert2} and NTv3. Evo2-7B was evaluated with the same activation-based detection procedure but did not yield an accepted candidate. Repository identifiers and checkpoint revisions were recorded for all measurements where available. For models originally evaluated without an explicit revision, the resolved Hub commit from the completed run is reported when recoverable; NTv3 remains an exception because its original model-weight revision was not recorded exactly.

The 22-model singleton causal cohort comprised this structural panel except Phi-3-mini. Phi-3 was represented structurally by a pre-frozen six-row basis rather than by one comparable primary row; Fig. 1B reports the median $q_1$ across those six rows. Supplementary Tables S1 and S2 describe the 22-model singleton causal cohort and therefore exclude Phi-3.

\subsection{Row-associated bilinear weight operator and structural metrics}

For output row $k$ of a gated feed-forward block, let $g_i$ and $u_i$ denote the gate- and up-projection row vectors for hidden unit $i$, and let $d_{k,i}$ denote the corresponding coefficient in row $k$ of the down projection. We defined the row-associated bilinear operator as

$U_k = \sum_i d_{k,i}\, g_i u_i^T.$

Its exact squared Frobenius norm can be evaluated without explicitly materializing all cross terms using the Gram identity

$\| U_k \|_F^2 = d_k^T K d_k,$

where

$K_{ij} = \left( g_i^T g_j \right)\left( u_i^T u_j \right),$

that is, $K$ is the Hadamard product of the gate- and up-projection Gram matrices. Singular values $\{\sigma_j\}$ were computed in float64. Spectral concentration was quantified as

$q_1 = \dfrac{\sigma_1^2}{\sum_j \sigma_j^2},$

Operator magnitude was quantified by

$\| U_k \|_F = \sqrt{\sum_j \sigma_j^2}.$

Thus, $q_1$ measures concentration of operator energy in the leading singular direction, whereas $\| U_k \|_F$ measures operator magnitude. For within-layer magnitude comparisons, layer-relative Frobenius magnitude was defined as the candidate's exact Frobenius norm divided by the median exact Frobenius norm of all rows in the same layer. A diagonal proxy that retains only per-hidden-unit terms omits interactions between distinct hidden units; it was used only for the retrospective Llama/Mistral/OLMo calibration in Fig. 1A and not for cross-model structural comparisons.

\subsection{Candidate provenance}

Candidate coordinates in the 22-model singleton causal cohort derive from two activation-based protocols. Sixteen primary candidates were selected under the activation-ratio detector, whereas six coordinates were frozen under an earlier activation-magnitude protocol and retained rather than reselected: Llama-7B, Mistral-7B, OLMo-7B, DNABERT-2, NTv3, and GENERator-EUK-3B. Phi-3 is not part of this 16+6 provenance breakdown; its six-row structural basis was frozen separately and is described below only in connection with its supplementary analysis.

For each layer $l$ and output row $k$, we defined $m_{lk} = \max_t \left| h_{ltk} \right|$, the maximum absolute down-projection output activation over all token positions $t$ in the single discovery forward pass. The activation-ratio score was $a_{lk} = m_{lk} / \mathrm{median}_j(m_{lj})$, where the median was taken across all output rows $j$ in the same layer. Coordinates selected under this protocol were accepted when $a_{lk} \geq 5.0$.

For Llama-7B, Mistral-7B, and OLMo-7B the coordinates are those published by Yu et al. and serve as calibration rather than as detector output; the ratio detector independently recovers the published coordinate in Llama-7B and Mistral-7B. In OLMo-7B it does not: the ratio maximum falls at L2/r269 and the activation maximum at L30/r269, both sharing the published coordinate's row index, so OLMo's headline coordinate is one instance of a depth-recurring row-269 family rather than a coordinate uniquely determined by the detector. These three coordinates were re-derived under the discovery protocol described below. The ratio maximum coincided with the published coordinate in Llama-7B (L2/r3968, ratio 3959.2, rank 1 of 131,072 layer--row coordinates) and Mistral-7B (L1/r2070, ratio 1714.3, rank 1 of 131,072); in OLMo-7B it fell at L2/r269 with the published coordinate at rank 2, and the absolute-activation maximum fell at L30/r269 as described. Repeating the measurement over 24 independent WikiText-2~\citep{merity2017pointer} windows rather than the single discovery input returned the same coordinate in every case, with the ratio maximum recovered in 100\% of windows for Llama-7B and Mistral-7B and 83\% for OLMo-7B, and was unchanged under both special-token conventions. Llama-7B shows the same depth-recurring pattern where its ratio maximum is L2/r3968 whereas its absolute-activation maximum is L30/r3968, the same row index at a much later depth. Mistral-7B is the exception, with both maxima at L1/r2070, so depth-recurring row families are the majority pattern among these three models rather than an OLMo-specific irregularity.

For DNABERT-2, the frozen coordinate L5/r603 is the global activation maximum under the canonical discovery preprocessing (out\_max = 944.5556, activation ratio 152.7345) but is not the global ratio maximum, which falls at L8/r603. The two rules disagree substantially here: the ratio-selected coordinate has $q_1$ = 0.379, below the minimum of the structural panel, while the frozen candidate has $q_1$ = 0.793. In NTv3 the two rules select L11/r1472 and L6/r1472 respectively, differing in layer but not in row index and in $q_1$ by less than 0.01. We retain the activation-based coordinates for these models and report the discrepancies rather than reconciling them retrospectively. Where the two rules disagree, the coordinate preferred by the ratio detector is not systematically the more consequential one. In OLMo-7B, ablating the published L1/r269 increased native loss by 111.8\% whereas ablating the ratio maximum L2/r269 increased it by 2.4\%, a 47-fold difference in favor of the coordinate that the ratio rule does not select, measured on the same evaluation pool with five seeded same-layer controls at L2 giving a median effect of $-2.5 \times 10^{-7}$. DNABERT-2 runs in the opposite direction. Ablating the ratio maximum L8/r603 raised masked-language-model loss by 10.7\% whereas ablating the frozen L5/r603 lowered it by 13.2\%. Neither rule is therefore uniformly closer to causal importance.

Detector execution for DNABERT-2 depends on tokenizer special-token handling: the historical discovery path uses default special tokens and reproduces the frozen activation maximum exactly, whereas the structural census path disables them and assigns the same coordinate an activation ratio near 1.0. Attention implementation has no effect on either result, and structural metrics for a given coordinate are identical across all four combinations, so the discrepancy is one of input preprocessing and not of model state.

Model checkpoints were revision-pinned where recorded. Resolved Hub commits are available for 21 of 22 models; NTv3's original model-weight revision was not recorded exactly and is not recoverable. Per-model provenance is given in Supplementary Table S1.

Text ratio-based discovery used the first 20 non-empty lines of the WikiText-2 raw test split, joined and tokenized with truncation to a maximum length of 512 tokens. Genomic discovery used the fixed 504-bp ACTB-derived sequence with model-specific tokenizer preprocessing. DNABERT-2 L5/r603 was defined by the maximum-activation discovery criterion using tokenizer-default special tokens; under this canonical preprocessing it is the global activation maximum (out\_max=944.5556, activation ratio 152.7345), although it is not the global ratio maximum. Calibration used batch size 1. Evo2-7B was evaluated with the ratio-based procedure and no candidate passed the predefined criterion (maximum ratio 2.22), so no structural spectrum was assigned.

To assess input stability without reselecting candidates, each frozen coordinate was additionally evaluated on 24 independent domain-matched inputs for the 20 models with a comparable singleton candidate and executable evaluation path. For each input, we recorded the candidate's maximum absolute down-projection output activation, its activation ratio relative to the same-layer median, its same-layer activation rank and percentile, and the position of the maximum activation where applicable. Candidate coordinates remained fixed throughout this assay; no causal measurements were used in selection or evaluation.

\subsection{Structural statistical analysis}

Within the 22-model singleton causal cohort, the random $q_1$ gap was defined for each model as candidate $q_1$ minus the mean $q_1$ of its five random same-layer controls. The top-norm $q_1$ margin was defined as candidate $q_1$ minus the maximum $q_1$ among the five highest-Frobenius-norm rows in the same layer after excluding the candidate in the same layer. The absolute structural margin used in structure--function analyses was the absolute value of this top-norm $q_1$ margin. To test whether a simple architecture/scale model explained cross-model spectral concentration, we fit the ordinary least-squares regression

$q_1 = \beta_0 + \beta_1 \log_{10}\left( N_{nonemb} \right) + \beta_2 I_{decoder} + \varepsilon$

across the 23 models with accepted candidates. Because the model panel is heterogeneous and not a random sample from a defined population of foundation models, coefficient tests are interpreted descriptively rather than as population-level architectural laws.

Correlations between structural descriptors and causal effect were computed on the full panel and, separately, within architecture class; panels with fewer than eight models are reported as underpowered and are not used to support any claim.

\subsection{Cross-model singleton functional-criticality census}

The functional-criticality census included 22 models with one frozen primary candidate per model: 10 text decoders, six text encoders, four genomic decoders, and two genomic encoders. Phi-3 was excluded because its frozen representation comprised six rows rather than one comparable singleton primary row; its separate tomography analysis is described in Supplementary Note S1. Candidate coordinates were frozen before causal evaluation and were not changed after observing causal responses. For each model, five distinct same-layer control rows were sampled without replacement using a deterministic SeedSequence(42) child stream assigned by panel index, excluding the frozen candidate set in that model/layer. Candidate and control rows were always evaluated one at a time.

Two control sets were constructed per model, both defined without reference to any causal measurement. Random same-layer controls were five distinct rows sampled without replacement from the candidate's layer using a deterministic SeedSequence(42) child stream assigned by panel index, excluding the frozen candidate. Top-norm same-layer controls were the five highest-exact-Frobenius-norm rows in the candidate's layer after excluding the frozen candidate, computed for all rows simultaneously via the Gram identity without materializing individual operators. The top-norm sweep used the same endpoints, window pools, masking realizations, batch constructions, and evaluation harness as the original census, differing only in which row indices were perturbed.

\subsubsection{Row intervention}

For a targeted down-projection row w, we cloned the original row, applied multiplicative suppression, evaluated the model, and restored the cloned row in a context-manager/finally path. The intervention was

$w_\varepsilon = (1-\varepsilon)\,w, \qquad \varepsilon \in \{0.5, 1.0\}.$

Thus $\varepsilon$=0.5 implemented 50\% row scaling and $\varepsilon$=1.0 set the targeted row to zero. Models were placed in evaluation mode and loaded/evaluated in float32; no autocast was declared. Candidate and control rows were never perturbed jointly.

\subsubsection{Native-objective evaluation endpoints}

Text decoders (n=10) were evaluated on WikiText-2-raw-v1 test text. Nonempty lines were shuffled with random.Random(42), concatenated, and tokenized separately for each model into 100 exactly 512-token windows. The endpoint was teacher-forced shifted-label mean token negative log-likelihood (NLL): logits at positions 0\ldots{}T-2 predicted labels at positions 1\ldots{}T-1, and summed NLL was divided by the number of predicted tokens. Models with more than 3B loaded parameters used batch size 4 (25 stored bootstrap units); smaller models used batch size 8 (13 units, including the final partial batch).

Text encoders (n=6) used the same WikiText construction but 256 windows of 512 tokens, evaluated in batches of 16. A single 15\% masking realization was generated with seed 42 and reused for baseline and every intervention condition. CLS, SEP, and PAD tokens were excluded from masking; selected tokens were replaced by MASK and all unselected labels were set to -100. The endpoint was summed cross-entropy over masked tokens divided by the masked-token count. MosaicBERT used the bert-base-uncased tokenizer; the remaining text encoders used their checkpoint tokenizers.

Genomic decoders (n=4) were evaluated on hg38 sequence windows derived from random\_262kb.bed. A seed-42 disjoint partition supplied 100 damage windows of 512 bp with less than 1\% N content. For GENERator models, the leftmost len mod 6 bases were trimmed, a BOS token was prepended, and tokenization used special tokens disabled; GenomeOcean used no 6-bp trim or forced BOS and also tokenized with special tokens disabled. The endpoint was teacher-forced shifted causal-LM mean token NLL, weighted over contributing tokens. Each genomic-decoder window was retained as an individual bootstrap unit. The separately constructed 96-window prompt pool was not used for this endpoint.

Genomic encoders (n=2) were evaluated on 256 hg38 windows of 600 bp sampled from random\_262kb.bed with random.Random(42), requiring less than 1\% N content. Inputs were padded/truncated to 256 tokens and evaluated in batches of 16. A fixed 15\% mask realization with seed 42 excluded special and padding tokens and was reused for baseline and every intervention. The endpoint was masked-nucleotide summed loss divided by the masked-token count. DNABERT-2 used the revision-pinned pretrained MLM checkpoint with the eager compatibility loader; NTv3 used its required pinned remote-code loader.

\subsubsection{Effect definitions and control comparison}

For evaluation unit j, the raw records stored summed loss Sj and the number of contributing tokens Nj. Baseline and perturbed losses were recomputed as token-weighted means:

$L_0 = (\sum_j S_{0,j}) / (\sum_j N_{0,j}), \qquad L_\varepsilon = (\sum_j S_{\varepsilon,j}) / (\sum_j N_{\varepsilon,j}).$

The signed relative effect was

$R_\varepsilon = (L_\varepsilon - L_0) / L_0,$

reported as 100$R_\varepsilon$ percent. Positive values therefore indicate degradation of the native objective after suppression, whereas negative values indicate a lower loss after suppression. For each model, the same-layer control reference was the median of the five independently measured control effects:

$G_\varepsilon = R_{\varepsilon,\mathrm{candidate}} - \mathrm{median}(R_{\varepsilon,\mathrm{control},1}, \ldots, R_{\varepsilon,\mathrm{control},5}).$

The 18/22 and 20/22 summaries count models with $G_\varepsilon$>0. The reported +0.60\% and +0.88\% cohort summaries are medians of $G_\varepsilon$ across the 22 models. Figure 2A-B display the signed candidate and individual-control $R_\varepsilon$ values, not $G_\varepsilon$. Figure 2C uses the signed full-ablation candidate R1.0 on the y-axis.

\subsubsection{Finite-intervention tomography}

Tomography followed the mechanistic-tomography framework of \citet{erramilli2026mechanistic}, which treats interventions as designed measurements, tests candidate observer families on held-out finite interventions at the intended scale, and expands the measurement family when structured residuals remain. We evaluated four observer families of increasing expressive capacity. F0 predicted a mask response by summing measured singleton effects. F1 applied a single least-squares calibration scalar to that sum. F2 jointly fit additive main effects by ridge regression. F3 augmented the design with all pairwise products $a_i a_j$ and fit main and pair coefficients jointly by ridge regression. Ridge strength was selected from a fixed grid on a calibration split; all reported performance was measured on a disjoint held-out split.

For DNABERT-2, the fixed basis contained 10 high-gain rows: (L5/r603), (L3/r86), (L3/r399), (L9/r264), (L9/r294), (L3/r603), (L3/r641), (L7/r603), (L6/r603), and (L5/r86). The lifted F3 design contained 10 main effects and 45 pair terms. Using seed 20260822, the mask pools comprised 78 fit masks, 20 calibration masks, and 20 held-out masks; the 78x55 lifted design had full column rank.

For each intervention strength, we reported held-out $R^2$, MAE, root-mean-square error, and normalized MAE. F2-to-F3 improvement was quantified as the relative reduction in held-out MAE. Confidence intervals were obtained from 5,000 paired bootstrap resamples of batch indices, resampling the held-out condition and its baseline jointly before baseline subtraction. Evidence for pair dependence was based on improved held-out prediction, bootstrap uncertainty, and identifiability of the lifted design rather than on training fit alone. A separate F0--F3 analysis of the fixed six-row Phi-3 basis is reported in Supplementary Note S1.

To assess sensitivity of the observer-family comparison to the specific fit/calibration/held-out partition, we performed 100 resplits per intervention strength, re-permuting role assignment among the fixed pool of 118 measured non-singleton mask conditions; pool sizes match exactly what was measured, so this resampling explores which subsets fall into held-out rather than new mask combinations. F3 outperformed F2 on held-out MAE in 100 of 100 splits at both $\varepsilon$=0.5 and $\varepsilon$=1.0. At $\varepsilon$=0.5, median [2.5, 97.5] held-out $R^2$ was $-$0.21 [$-$0.45, 0.06] for F0, 0.30 [$-$0.35, 0.57] for F1, 0.55 [0.26, 0.70] for F2, and 0.92 [0.79, 0.96] for F3; at $\varepsilon$=1.0 the corresponding values were $-$0.45 [$-$1.14, $-$0.05], 0.31 [$-$0.33, 0.69], 0.56 [0.33, 0.70], and 0.76 [0.49, 0.88]. The selected ridge $\lambda$ was unstable across resplits --- F2's selected $\lambda$ ranged from 0.001 to 100 at $\varepsilon$=0.5 and 0.001 to 30 at $\varepsilon$=1.0, and F3's from 0.001 to 10 at $\varepsilon$=0.5 and 0.1 to 100 at $\varepsilon$=1.0 --- indicating that the point estimate depends on which conditions fall into the fit set, although the qualitative F3 > F2 ranking did not.

\subsubsection{DNABERT-2 pretrained masked-language-model experiment}

DNABERT-2 was evaluated with the revision-pinned pretrained masked-language-model checkpoint using eager attention. Sequence windows were sampled from the hg38 reference genome using random\_262kb.bed. We evaluated 256 windows of 600 bp, in 16 batches, with a fixed 15\% mask realization (seed 42), yielding 4,453 masked tokens. The same masks were reused across all intervention conditions. Pair epistasis was defined as the joint loss change minus the sum of the two singleton loss changes. The L9/r264 $\times$ L9/r294 epistasis values were obtained in a standalone direct pair-ablation experiment in which the two rows were perturbed individually and jointly. This direct joint-ablation condition was separate from the F0--F3 tomography pools and does not appear among the 118 measured non-singleton mask conditions used for fitting, calibration, and held-out evaluation.

For DNABERT-2, candidate selection and causal evaluation were distinct assays. L5/r603 was defined using the fixed 504-bp ACTB discovery probe tokenized with the model's default special tokens; this configuration produces the characteristic activation maximum at that coordinate (out\_max=944.5556; activation ratio 152.7345; activation rank 1). Functional evaluation instead used the independent seed-42 hg38 masked-language-model dataset described above. The ACTB discovery preprocessing and hg38 causal-evaluation preprocessing were therefore distinct.

\subsubsection{GENERator GC-composition and BOS analyses}

For the GENERator composition-specificity analysis, two disjoint hg38 window pools were used: 96 windows of 170 bp for generation, with the first 120 bp used as prompts, and 100 windows of 512 bp for teacher-forced native-loss measurement. The pools were drawn from non-overlapping regions of the same reference BED file. Native-loss damage for row 2371 was measured over row scales $\alpha \in \{0,0.25,0.5,0.75,1.0,1.5,2.0,3.0,5.0\}$, with the $\alpha$=0 ablation defining the target damage. Five seeded non-2371 control rows were each swept over $\alpha \in \{0,0.1,\ldots,8.0\}$ to identify the closest achievable damage. A second specificity control replaced the weight vector at the row-2371 location with one fixed random unit direction, sampled once and reused unchanged at every tested scale $c \in \{0.0125, 0.5, 1.0, 3.0, 8.0\}$. Because the row is replaced rather than perturbed additively, c=0 is equivalent to ablation and native-loss damage decreases monotonically with increasing c; the maximum tested scale, eight times the row's own norm, still retained 84\% of the ablation damage. Generated-sequence statistics were evaluated at each c over 96 prompts and three generation seeds, with per-prompt GC retained. GC fraction was computed over the generated continuation after filtering to A/C/G/T characters, and 95\% confidence intervals were obtained from 5,000 bootstrap resamples paired over prompts.

For the BOS attention/activation reproduction, 40 seed-42 hg38 windows of 512 bp were tokenized directly with tokenizer-default special-token handling, without the 6-bp trimming used for the native-loss endpoint and without manually prepending BOS; inputs were truncated to a maximum of 128 tokens. We measured mean incoming attention mass assigned to position 0, the corresponding fold over the uniform expectation, the fraction of layer-head observations whose attention maximum occurred at position 0, and the position of the maximum high-gain activation. For each window of tokenized length $L$, the uniform-attention expectation was $1/L$; the reported fold enrichment was the mean incoming attention mass at position 0 divided by the mean of these per-window uniform expectations. Because position 0 has special status in a causal decoder and no independent causal manipulation of the attention sink was performed, this analysis is interpreted as co-occurrence only.

For the position-specific mediation experiment, we intervened on the contribution of GENERator-EUK row L4/r2371 at selected sequence positions while leaving the underlying model weights fixed. Six conditions were evaluated: intact model; full row ablation; removal of the row contribution only at BOS; removal of the row contribution at all positions except BOS; full row ablation with restoration of the intact BOS contribution; and full row ablation with restoration of the corresponding contribution at a matched non-BOS position. Native NLL was evaluated on the same 100-window damage pool used for the existing GENERator intervention analyses. Generated GC fraction was evaluated on the generation prompt pool using the same decoding procedure. Rescue fraction was defined relative to intact and full-ablation effects as $(E_{ablation} - E_{intervention}) / (E_{ablation} - E_{intact})$, with paired 95\% bootstrap intervals computed over the stored evaluation units. During autoregressive generation, position-specific intervention was applied during prefill and was disabled for subsequent KV-cached decoding steps when the target position was no longer present.

\subsubsection{Statistical analysis}

Exact operator spectra are deterministic single-checkpoint calculations and therefore are shown without sampling error bars. Tomography confidence intervals use 5,000 paired bootstrap resamples over batch indices, as described above. The manuscript reports effect sizes and held-out predictive performance rather than treating cross-model comparisons as a formal population test, because the model panel is small and heterogeneous. No multiple-comparison correction was applied across predictors, intervention strengths, or model subgroups.

Per-condition 95\% confidence intervals were obtained from 5,000 percentile bootstrap draws over the stored evaluation units. In each draw, the same resampled index vector selected the paired baseline and perturbed units, each loss was recomputed as total summed loss divided by total contributing tokens, and $R_\varepsilon$ was recalculated. The stored units were actual batches for text decoders and both encoder groups, and individual windows for genomic decoders. Condition-specific random streams were drawn sequentially from SeedSequence(42).spawn(1000) in frozen panel/condition order.

For the cohort candidate-minus-control gap, the 22 model-level $G_\varepsilon$ values were resampled with replacement and the median was recomputed for each of 5,000 draws. Percentile intervals used seed 47 for $\varepsilon$=0.5 and seed 52 for $\varepsilon$=1.0. To test the relationship between structural concentration and the full-ablation functional effect, Spearman's rank correlation was computed between candidate $q_1$ and the signed full-ablation candidate effect across the 22 models. The reported p value is SciPy's default asymptotic two-sided test. A 5,000-draw model bootstrap resampled the 22 model rows with replacement, recomputed $\rho$, discarded only nonfinite replicates, and formed percentile limits (seed 43). Negative and zero effects were retained without transformation.

\subsubsection{Data and code availability}

\ifdefined\TMLRArxiv
Code and non-model artifacts are available at https://github.com/Georgakopoulos-Soares-lab/superweights. The repository contains all analysis code, per-model structural and causal measurement scripts, and result artifacts for every experiment reported here, with two exceptions inherent to their size: the hg38 reference FASTA and WikiText-2 are not committed and must be obtained independently. Reproducing the full 22-model census additionally requires downloading the pinned Hugging Face checkpoint revisions listed in Supplementary Table S1.
\else
Code and non-model artifacts are provided in the anonymized supplementary material. The supplementary material contains all analysis code, per-model structural and causal measurement scripts, and result artifacts for every experiment reported here, with two exceptions inherent to their size: the hg38 reference FASTA and WikiText-2 are not included and must be obtained independently. Reproducing the full 22-model census additionally requires downloading the pinned Hugging Face checkpoint revisions listed in Supplementary Table S1.
\fi

\ifdefined\TMLRArxiv
\subsubsection*{Author Contributions}

A.T., A.K., and I.G.S. jointly conceived the study. A.T. and A.K. designed the methodology, developed the software, performed the experiments and formal analyses, and generated the visualizations. A.T. and A.K. interpreted the results and wrote the original manuscript. I.G.S. provided supervision, contributed to interpretation of the results, and reviewed and edited the manuscript. I.G.S. acquired funding for the project. All authors reviewed and approved the final manuscript.

\subsubsection*{Acknowledgments}

Research reported in this publication was supported by the National Institute of General Medical Sciences of the National Institutes of Health under award number R35GM155468.
\fi

\subsubsection*{Competing interests}

The authors declare no competing interests.

\bibliography{references}
\bibliographystyle{tmlr}

\clearpage
\appendix
\section*{Supplementary material}

Supplementary Figures S1--S2 and Supplementary Note S1 are reproduced below.

\begin{center}
\includegraphics[width=0.8\linewidth]{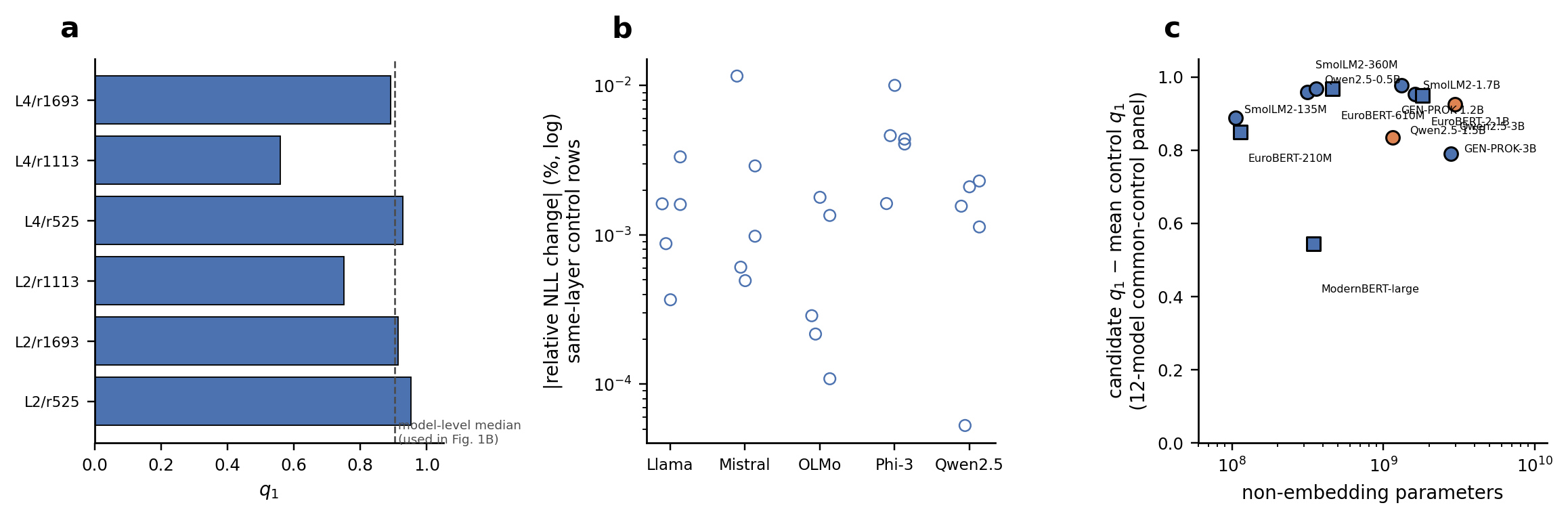}

\textbf{Supplementary Figure S1. Structural and causal control detail.} \textbf{(A)} Exact $q_1$ values for all six Phi-3 candidate rows. \textbf{(B)} Individual same-layer random-control causal effects for all five decoder models in the E10 causal audit. \textbf{(C)} Candidate-minus-mean-random-control $q_1$ gap across the 22-model causal cohort versus non-embedding parameter count.
\end{center}

\textbf{Supplementary Note S1. Phi-3 finite-intervention tomography.} The F0--F3 observer-family progression was also applied to the pre-frozen six-row Phi-3 basis shown in Supplementary Fig. S1A. No third-order model was fit. The full-ablation response is therefore treated as unresolved rather than modeled with higher-order terms after observing the data, and the Phi-3 tomography result is not used to support any main-text cohort claim.

\begin{center}
\includegraphics[width=0.8\linewidth]{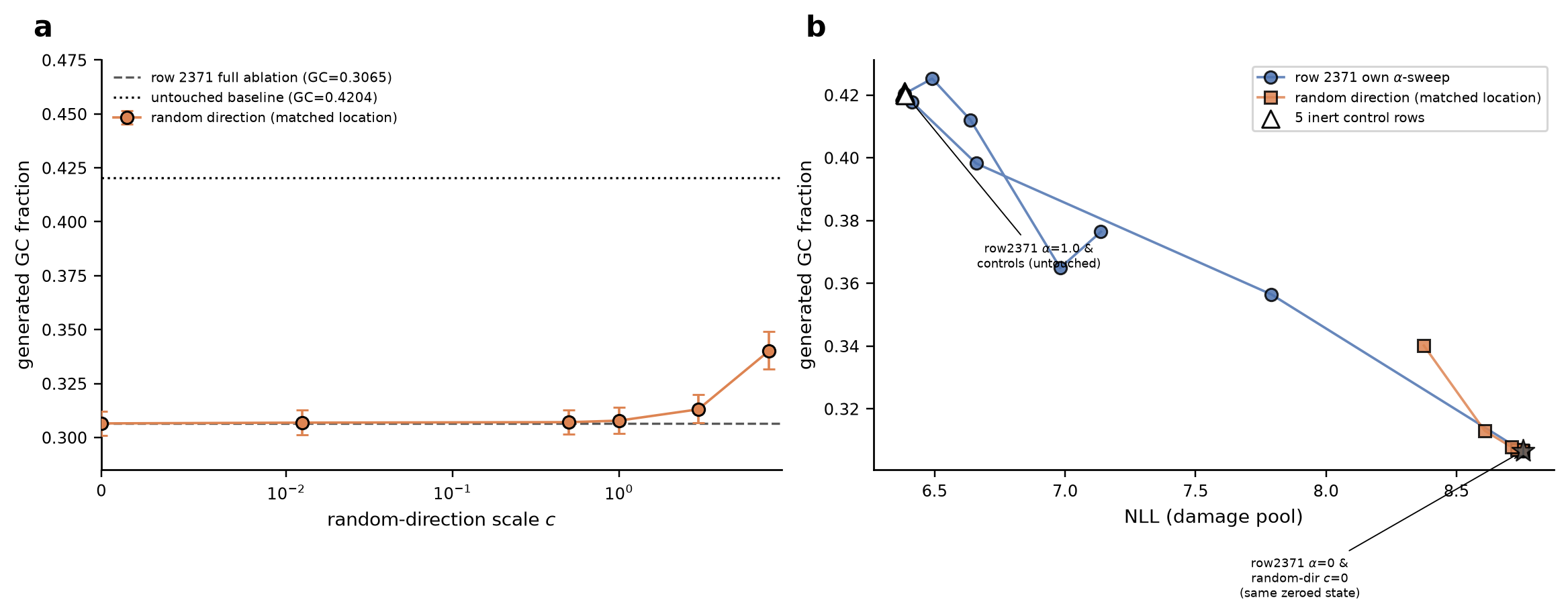}

\textbf{Supplementary Figure S2. Damage-matched random-direction analysis of GENERator EUK row L4/r2371.} Left: generated GC fraction across scales of one fixed random unit direction replacing the learned row-2371 weight vector. Right: generated GC fraction versus native-loss damage for the row-2371 scale sweep, five non-2371 control rows, and the fixed random-direction replacement sweep. The experiment tests whether the compositional phenotype requires preservation of the learned row direction; even the largest random replacement remains substantially damaging.
\end{center}

\end{document}